\documentclass{article}

\usepackage{arxiv}

\usepackage[utf8]{inputenc} 
\usepackage[T1]{fontenc}    
\usepackage{hyperref}       
\usepackage{url}            
\usepackage{booktabs}       
\usepackage{amsfonts}       
\usepackage{nicefrac}       
\usepackage{microtype}      
\usepackage[pdftex]{graphicx}
\usepackage{caption}
\usepackage{amsmath}

\title{Optimal Network Robustness Against Attacks\\ in Varying Degree Distributions}

\author{
  Masaki Chujyo\thanks{Corresponding author} \\
  Japan Advanced Institute of Science and Technology\\
  Ishikawa, 923-1292 Japan\\
  \texttt{mchujyo@jaist.ac.jp} \\
   \And
 Yukio Hayashi \\
  Japan Advanced Institute of Science and Technology\\
  Ishikawa, 923-1292 Japan\\
  \And
  Takehisa Hasegawa \\
  Graduate School of Science and Engineering, Ibaraki University, \\
  Ibaraki, 310-8512 Japan \\
}

\begin{document}
\maketitle

\begin{abstract}
In varying degree distributions, we investigate the optimally robust networks against targeted attacks to nodes with higher degrees.
In considering that a network tends to have more robustness with a smaller variance of degree distributions, we clarify the optimal robustness at random regular graphs in their comprehensive discrete or random perturbations.
By comparing robustness measurements on them, we find that random regular graphs have the optimal robustness against attacks in varying degree distributions. 
\end{abstract}


\section{Introduction}
At the beginning of this century, it has been found that many real-world networks commonly have a scale-free structure in which the degree distribution follows a power-law \cite{Barabasi1999}. 
Unfortunately, such networks are extremely vulnerable to targeted removals (attacks) of nodes with higher degrees \cite{albert2000error}.
However, our modern society is supported by scale-free networks, such as the Internet, communication networks, traffic systems, power grids, social networks, supply chain networks, protein-protein interaction networks, and metabolic networks.
Therefore, overcoming the vulnerability of real-world networks has been an important issue.

We focus on varying the degree distribution for investigating the robustness of connectivity.
As well-known results, scale-free networks with power-law degree distributions are more vulnerable to attacks than Erd\H{o}s-R\'enyi random graphs with exponential degree distributions, while scale-free networks are more tolerant to random node removals (failures) than Erd\H{o}s-R\'enyi random graphs \cite{albert2000error}.
Recently, it has been shown that random networks with smaller variances of degree distributions are more robust through numerical simulations in the range of power-law, exponential, and narrower degree distributions \cite{chujyo_CN}.
The continuously varying degree distributions are generated by the growing network model \cite{Krapivsky2000, Krapivsky2001, Krapivsky2002} and the inverse preferential model \cite{Liao}. 
In addition, for discussing the pure effect of degree distributions on the robustness, the networks are randomized thought configuration models \cite{config_1,config_2}.
In a special class of networks with multimodal distributions including power-law ones, bimodal networks with two kinds of degrees are the most robust for maximizing the sum of two critical thresholds of whole fragmentations by both failures and attacks \cite{tanizawa2006opt}.
In other words, the robustness against both failures and attacks increases, as the variance of degree distribution decreases. 
At pinpoints, in comparing the robustness of regular graphs, Erd\H{o}s-R\'enyi random graphs \cite{erdos1959random}, Watts-Strogatz models \cite{watts1998collective}, and Barab\'asi Albert models \cite{Barabasi1999}, it has been numerically shown that regular graphs are the most robust \cite{ma2015theoretical}.
Moreover, the degree distribution becomes narrower in maximizing the robustness index by random rewiring \cite{ma2015theoretical}.
Although the maximization changes the degree distribution to trimodal or tetramodal distributions with three or four degrees, this is not enough to say that regular graph with one degree is the most robust.

On the other hand, it is found that an onion-like network with positive degree-degree correlations \cite{Newman2002} is the most robust under fixing a degree distribution \cite{Schneider2011, tanizawa2006opt}. 
By increasing the degree-degree correlations, some rewiring methods have been proposed for improving the robustness \cite{Xulvi2004, Wu2011}.
In addition, an incrementally growing method for constructing onion-like networks is proposed by enhancing interwoven long loops \cite{hayashi2018new, hayashi2018onion}.
The robust networks generated by this growing method have exponential degree distributions \cite{hayashi2018new}.
This result also suggest that more homogeneous degree distributions are crucial to increase the robustness.

Besides the degree-degree correlations, loops on networks have been getting attention for increasing the robustness \cite{Braunstein2016,hayashi2018onion,chujyo2021loop}.
The relation between loops and the robustness is supported by an asymptotically equivalence of network decycling and network dismantling \cite{Braunstein2016}.
Network decycling or feedback vertex set (FVS) is a minimum set of nodes that removal makes the network without loops, while network dismantling is a minimum set of nodes whose removal makes it a smaller size of connected components.
Intuitively, networks without loops are easily fragmented by any node removals. 
As the importance of loops, it has numerically shown that networks with a larger size of FVS have more robustness against attacks in the incrementally growing onion-like networks \cite{hayashi2018onion}.
Furthermore, loop-enhancing rewiring methods by increasing the size of FVS have been proposed \cite{chujyo2021loop}. 
When the loop-enhancing rewiring is applied to a network for increasing the robustness, the variance of degree distributions becomes smaller. 
It also shows that decreasing of the variance of degree distribution is strongly related to increasing the robustness.

These previous studies suggest that a network tends to be more robust against attacks with a smaller variance of degree distribution.
Thus, a random regular graph with the minimum (zero) variance of degree distribution is predicted to have the optimal robustness.
In this paper, we clarify the optimal robustness in varying degree distributions by comparing the robustness of random regular graphs and their perturbed ones by comprehensive discrete or random perturbations.

\section{Surrounding of random regular graphs}
A regular graph consists of all nodes with a constant degree. 
Thus, the variance of degree distributions is zero.
We compare the robustness in random regular graphs and the perturbed ones around it.
As the surroundings, we consider two types of networks with discrete and random perturbations.
In Sec. 2.1 for discrete perturbations, we introduce bimodal networks with two types of degrees whose modality is the second minimum to the regular graph with only one degree.
In Sec. 2.2 for random perturbations which include several modalities of degrees, we introduce modified networks by adding and removing links to the regular graphs uniformly at random.

\subsection{Discrete perturbations}
As discrete perturbations of a random regular graph, we introduce bimodal networks with two degrees $d_1<d_2$ under the average degree $d$.
For the bimodal networks, there are several combinations of degrees $d_1$ and $d_2$ in $\Delta d=d_2-d_1 \ge 2$.

For given degrees $d_1$ and $d_2$ for a bimodal network with $N$ nodes and the average degree $d$, the number of nodes $N_1$ and $N_2$ corresponding to degrees $d_1$ and $d_2$ are derived as follows. 
From the total number of nodes $N$ and links $M$, 
\begin{equation}
    N=N_1+N_2 ,
\end{equation}
\begin{equation}
    M=\frac{d\times N}{2}=\frac{d_1\times N_1}{2}+\frac{d_2 \times N_2}{2} ,
\end{equation}
we obtain
\begin{equation}
    N_1=\frac{d_2-d}{\Delta d}N ,
    \label{N1}
\end{equation}
\begin{equation}
    N_2=\frac{d-d_1}{\Delta d}N .
    \label{N2}
\end{equation}
Since $N_1$ and $N_2$ must be positive integers, $N$ needs to be divisible by $\Delta d$.
As shown in Table \ref{tab_bimodal}, the combinations are $(d_1, d_2)=(d-1, d+\Delta d-1), (d-2, d+\Delta d-2), ..., (d-\Delta d+1, d+1)$.
Except for a star structure with $d_1=1$ and $d_2=N-1$ which is obviously vulnerable to attacks, Table \ref{tab_bimodal} shows all possible combinations in the ranges of $2\le d_1\le d-1$ and $d+1\le d_2 \le N-2$ for constant $N$ and $d$. 
Note that these combinations of degrees are comprehensive around a regular graph.
For constructing the bimodal networks, we use a configuration model according to the degree distribution:
\begin{equation}
P(d_1)=N_1/N, 
\label{PK_BimP-1}
\end{equation}
\begin{equation}
P(d_2)=N_2/N. 
\label{PK_BimP-2}
\end{equation}
After randomizing them through the configuration model \cite{config_1, config_2}, we can discuss the pure effect of degree distributions on the robustness. 
\begin{table}
    \caption{Two degrees $d_1$ and $d_2$, the variance $\sigma^2$ of degrees, and the fraction $N_1/N$ of nodes with degree $d_1$ in bimodal networks with the average degree $d=6$.}
    \label{tab_bimodal}
    \centering
    \begin{tabular}{ccccc}
    \hline \hline
        $\Delta d$&  $d_1$& $d_2$&  $\sigma^2$&  $N_1/N$ \\ \hline
         2&   5& 7&  1&  1/2 \\ \hline
         3&   5& 8&  2&  2/3 \\
          &   4& 7&  2&  1/3 \\ \hline
         4&   5& 9&  3&  3/4 \\
         &   4& 8&  4&  2/4 \\
         &   3& 7&  3&  1/4 \\\hline
         5&   5& 10&  4&  4/5 \\
         &   4& 9&  6&  3/5 \\
         &   3& 8&  6&  2/5 \\
         &   2& 7&  4&  1/5 \\ \hline
         &$\vdots$   & &$\vdots$  & \\ \hline
         $\Delta d$& $d-1=5$ & $d+\Delta d-1$ &  $\Delta d-1$ & $(\Delta d-1)/\Delta d$ \\
          & $d-2=4$ & $d+\Delta d-2$ &  $2(\Delta d-2)$ & $(\Delta d-2)/\Delta d$ \\
          & $d-3=3$ & $d+\Delta d-3$ &  $3(\Delta d-3)$ & $(\Delta d-3)/\Delta d$ \\
          & $d-4=2$ & $d+\Delta d-4$ &  $4(\Delta d-4)$ & $(\Delta d-4)/\Delta d$ \\
        \hline \hline
    \end{tabular}
\end{table}

\begin{figure}
    \centering
    \includegraphics[scale=0.65]{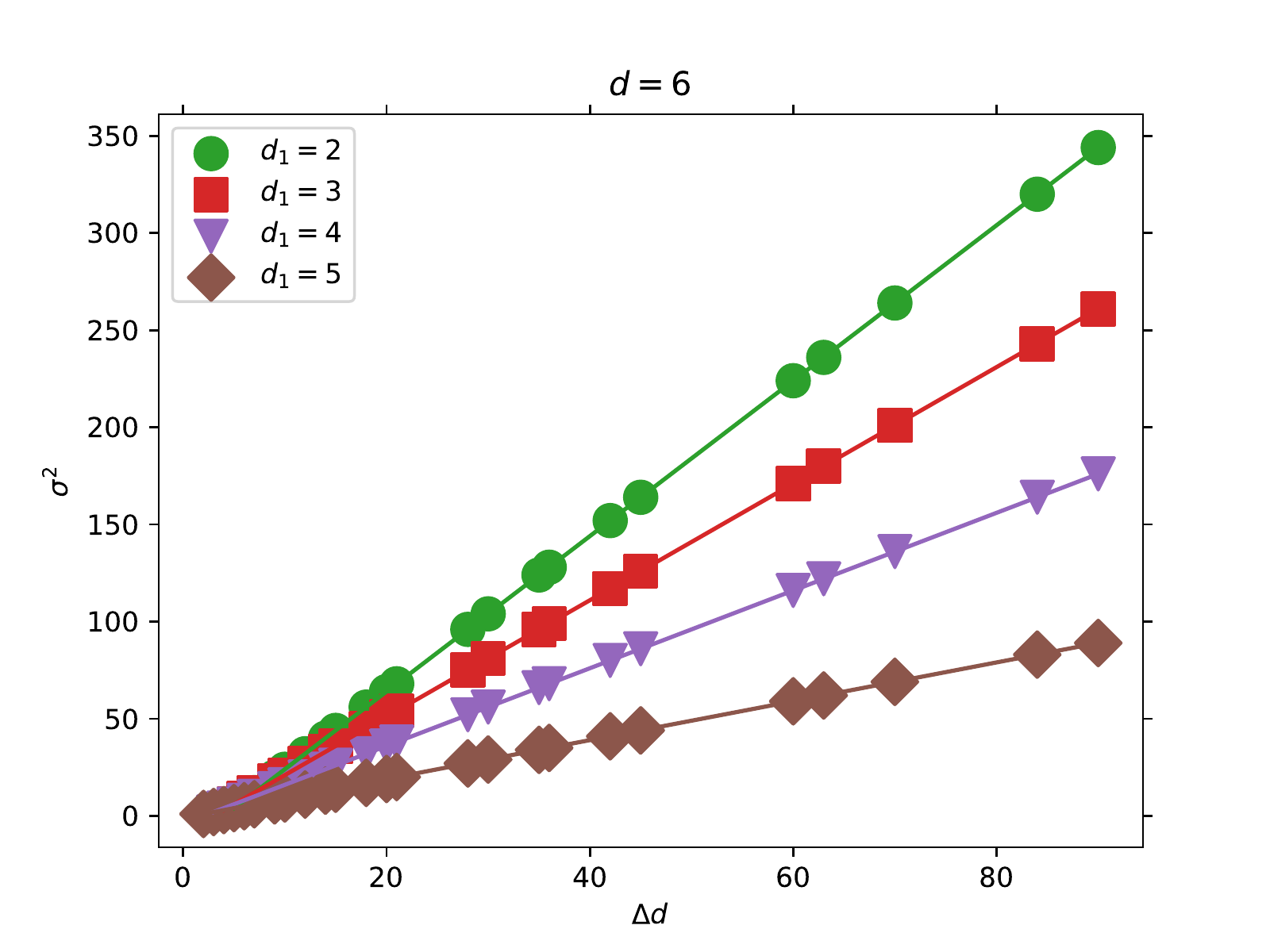}
    \caption{Variances $\sigma^2$ of degree distributions versus $\Delta d$ in bimodal networks for $N=1260$ and $d=6$. Green, red, violet, and brown lines with circle, square, inverted triangle, and diamond points show the results for $d_1=2$, $3$, $4$, and $5$, respectively. Each line is straight, since $\sigma^2$ and $\Delta d$ are proportional for a constant $d_1$ in Eq. (\ref{prop_delta_d_simga}).}
    \label{fig_bimodal_variance_delta}
\end{figure}

We can easily calculate the variance of degree distribution of a bimodal network.
From Eqs. (\ref{N1}) and (\ref{N2}), the variance is derived as follows: 
\begin{eqnarray}
    \sigma^2 &=& \langle k^2 \rangle -\langle k \rangle^2 \nonumber \\
    &=& \frac{1}{N} (d_1^2 \times N_1+d_2^2 \times N_2)-d^2 \nonumber \\
    &=& \frac{1}{N} \left( d_1^2 \frac{d_2-d}{\Delta d} N + d_2^2 \frac{d-d_1}{\Delta d}N \right)-d^2 \nonumber \\
    &=& \frac{d_1^2(d_2-d)+d_2^2(d-d_1)}{d_2-d_1}-d^2 \nonumber \\
    &=& (d_2-d)(d-d_1) .
    \label{variances}
\end{eqnarray}
When each of $d_1$ or $d_2$ is close to $d$, the variance $\sigma^2$ is small. 
In particular, the minimum variance is $\sigma^2=1$ at $d_1=d-1$ and $d_2=d+1$. 
As shown in Fig. \ref{fig_bimodal_variance_delta}, the variances $\sigma^2$ and $\Delta d$ are proportional for a constant $d_1$.
By substituting $\Delta d=d_2-d_1$ into Eq. (\ref{variances}), we obtain 
\begin{eqnarray}
    \sigma^2 &=&  (d_2-d)(d-d_1) \nonumber \\
      &=&(d_1+\Delta d-d)(d-d_1) \nonumber \\
      &=&(d-d_1)\Delta d-(d-d_1)^2 .
    \label{prop_delta_d_simga}
\end{eqnarray}
The increasing of the variances $\sigma^2$ is remarkable for smaller $d_1$, e.g. green or res lines in Fig. \ref{fig_bimodal_variance_delta}.
Note that $\Delta d$ is a divisor of $N$ from Eqs. (\ref{N1})(\ref{N2}).

\subsection{Random perturbations}
\begin{figure}
    \centering
    \includegraphics[scale=0.65]{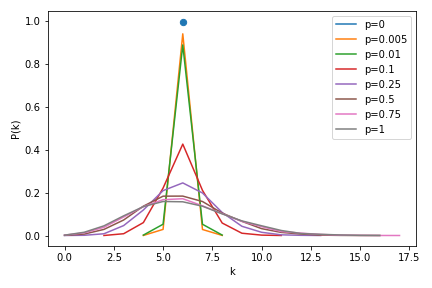}
    \caption{Degree distributions of random perturbations to random regular graphs with $N=10000$ and $d=6$. The degree distributions change from a delta function of regular graphs ($p=0$) to  Poisson distributions ($p=1$). }
    \label{Pk_RP}
\end{figure}

As random perturbations of regular graphs, we introduce modified networks by adding and removing links to random regular graphs. 
Here, $0\le p\le 1$ is the ratio of removed links to the existing links.
After removing, for fixing the average degree $d$, the same number of $Mp$ links are added at randomly chosen nodes in prohibiting multi-links and self-loops.
At $p=0$, all links are unchanged, while at $p=1$, all links are rewired as Erd\H{o}s-R\'enyi random graphs \cite{erdos1959random}.
For $0<p<1$, we can derive the degree distribution as follow.
\begin{equation}
    P(k) = \sum_{k_1 + k_2 = k} \binom{d}{k_1} (1-p)^{k_1} p^{d-k_1}  \frac{\lambda^{k_2}}{k_2!} e^{-\lambda} ,
    \label{PK_randomP}
\end{equation}
where $k_1$ and $k_2$ are the number of unremoved and added links to a node, respectively, and $\lambda = 2 d p$. 
Fig \ref{Pk_RP} shows 100 averaged degree distributions for each $p$.
As $p$ increases, the degree distributions become wider from a delta function $P(k)=\delta_{k,d}$ of the regular graph to Poisson distribution. 
Even for $p=0.005$ and $0.01$, the degree distribution consists of five kinds of degrees, whose modality is much more than the trimodal or tetramodal distributions as in \cite{ma2015theoretical}.

\section{Robustness analysis}
In this section, we consider two measurements of the robustness index and the percolation threshold for comparing the robustness against attacks in regular graphs and perturbed ones.
The robustness index \cite{Schneider2011} is defined as 
\begin{equation}
R_{\mathrm{TA}} \stackrel{\rm{def}}{=} \frac{1}{N}\sum_{q=1}^N S(q) ,
\label{R_TA}
\end{equation}
where $q$ is the number of removed nodes by attacks, and $S(q)$ is the fraction of nodes in the largest connected components.
As attacks, nodes are removed one-by-one from the node with the highest degree.
Note that the ranges of $R_{\mathrm{TA}}$ are $[1/N, 0.5]$.

The percolation threshold $f_c$ is the fraction of remaining nodes (occupied nodes in percolation theory), where whole networks are fragmented.
Below $f_c$, there are no the largest connected components for the network in the thermodynamic limit ($N \rightarrow \infty$).
As $f_c$ decreases, the network becomes more robust, because more nodes need to be removed for fragmentation.
We describe the method to estimate the percolation threshold $f_c$ against attacks by using generating functions for any degree distributions \cite{newman2001random, cohen2001breakdown}. 
Remember that $P(k)$ is the degree distributions.
Here, we define the excess degree distribution $Q(k)=(k+1)P(k+1)/\langle k \rangle$, which means that the probability of reaching a node with $k$ by following a randomly selected link.
We consider the attacks that remove $1-f$ fraction of nodes in order of higher degrees.
Then, $k_\mathrm{cut}$ denotes the highest degree in the remaining nodes, and $\Delta f$ denotes the fraction of removed nodes with degree $k_\mathrm{cut}$.
The probability that randomly selected node is not removed is 
\begin{equation}
    f=\sum_{k=k_{\mathrm{min}}}^{k_\mathrm{cut}} P(k)-\Delta fP(k_\mathrm{cut}) .
    \label{f_define}
\end{equation}
The probability that a node by following a randomly selected link is not removed is 
\begin{equation}
    \hat{f}=\sum_{k=k_{\mathrm{min}}}^{k_\mathrm{cut}} Q(k-1)-\Delta fQ(k_\mathrm{cut}-1) .
\end{equation}
Let $\hat{P}(k)$ be the degree distribution of the network remaining after the attacks, and $\hat{Q}(k)$ be the excess degree distribution. 
By following \cite{cohen2000resilience, cohen2001breakdown}, we can derive 
\begin{eqnarray}
    \hat{P}(k)=\frac{1}{f} \left[ \sum_{k'=k_\mathrm{min}}^{k_\mathrm{cut}}P(k') \binom{k'}{k}\hat{f}^k (1-\hat{f})^{k'-k} -\Delta f P(k_\mathrm{cut}) \binom{k_\mathrm{cut}}{k}\hat{f}^k (1-\hat{f})^{k_\mathrm{cut}-k}\right], \\
    \hat{Q}(k)=\frac{1}{\hat{f}} \left[ \sum_{k'=k_{\mathrm{min}}}^{k_\mathrm{cut}}Q(k') \binom{k'}{k}\hat{f}^k (1-\hat{f})^{k'-k} -\Delta f Q(k_\mathrm{cut}) \binom{k_\mathrm{cut}-1}{k}\hat{f}^k (1-\hat{f})^{k_\mathrm{cut}-1-k} \right].
\end{eqnarray}
We consider the generating functions $F_0(x)$ and $F_1(x)$ for the distributions $\hat{P(k)}$ and $\hat{Q(k)}$ after attacks, 
\begin{eqnarray}
    F_0(x)=\sum_k \hat{P}(k)x^k=\frac{1}{f} \sum_{k'=k_\mathrm{min}}^{k_\mathrm{cut}}P(k') (\hat{f}x+1-\hat{f})^{k'} -\frac{\Delta f}{f} P(k_\mathrm{cut}) (\hat{f}x+1-\hat{f})^{k_\mathrm{cut}}, \\
    F_1(x)=\sum_k \hat{Q}(k)x^k=\frac{1}{\hat{f}} \sum_{k'=k_\mathrm{min}-1}^{k_\mathrm{cut}-1}Q(k') (\hat{f}x+1-\hat{f})^{k'} -\frac{\Delta f}{\hat{f}} Q(k_\mathrm{cut}-1) (\hat{f}x+1-\hat{f})^{k_\mathrm{cut}-1}. 
\end{eqnarray}
The fractions of nodes belonging to the largest connected components after attacks is 
\begin{equation}
    s=f(1-F_0(u)) ,
    \label{s_is_here}
\end{equation}
where $u$ is the solution of the self-consistent equation of $u=F_1(u)$.
We numerically obtain the fixed point $u$.
Here, $s$ is equivalent to $S(q)$ for $q=1-f$ in Eq. (\ref{R_TA}).
Since the existence of $u < 1$ solution is a condition for the appearance of a largest connected component ($s > 0$), the percolation threshold $f_c$ can be obtained under the following conditions, 
\begin{eqnarray}
    F'_1(1) &=& \sum_{k'=k_\mathrm{min}-1}^{k_\mathrm{cut}-1}k'Q(k') \nonumber \\ 
    &=& \sum_{k=k_\mathrm{min}}^{k_\mathrm{cut}} \frac{k(k-1)P(k)}{\langle k \rangle}-\Delta f \frac{k_\mathrm{cut}({k_\mathrm{cut}-1})P({k_\mathrm{cut}})}{\langle k \rangle}=1.
    \label{fc_define}
\end{eqnarray}
From Eq. (\ref{f_define}), we numerically obtain $f_c$ from Eq. \ref{fc_define} with $k_\mathrm{cut}$ which satisfies the condition of Eq. (\ref{fc_define}).
Note that $f_c$ can be only applied on randomized networks like the configuration model.
Here, we apply Eqs. (\ref{f_define}-\ref{fc_define}) for the degree distributions of Eqs. (\ref{PK_BimP-1}, \ref{PK_BimP-2}, \ref{PK_randomP}) in bimodal networks and randomly perturbed networks from the regular graphs. 
random regular graphs and their perturbations.

\section{Results}
We compare these measurements $R_{\mathrm{TA}}$ and $1-f_c$ for the robustness against attacks in random regular graphs and perturbed ones with $N=6300$ nodes and the average degree $\langle k \rangle=4$ or 6.
Note that for $N=6300$, it is possible to generate regular graphs for $d=4$ and $6$. 
We use bimodal networks of all possible combinations of positive integers $d_1$ and $d_2$ with $2\le \Delta d \le 10$, as shown in Table \ref{tab_bimodal}. 
In addition, we use modified networks by random perturbations for the ratios of $p=0.005, 0.01, 0.1, 0.25, 0.5, 0.75,$ and $1$.

\begin{figure}[htp]
    \begin{minipage}[b]{0.5\textwidth}
        \includegraphics[width=\linewidth]{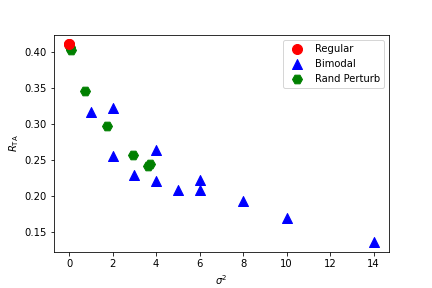}
        \caption*{(A) For $\langle k \rangle=4$.}
    \end{minipage}  
    \begin{minipage}[b]{0.5\textwidth}
        \includegraphics[width=\linewidth]{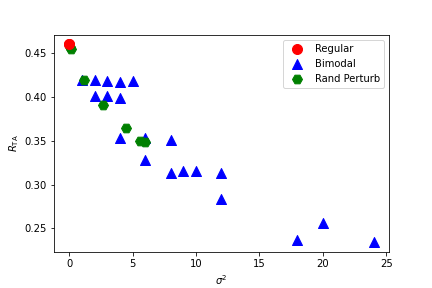}
        \caption*{(B) For $\langle k \rangle=6$.}
    \end{minipage}
    \caption{The robustness index $R_{\mathrm{TA}}$ versus the variances $\sigma^2$ of degree distributions. The results for for the average degree \textbf{(A)} $\langle k \rangle=4$, and \textbf{(B)} $\langle k \rangle=6$. In both figures, $R_{\mathrm{TA}}$ becomes higher as the variance decreases. In particular, the random regular graphs (red points) are the most robust.}
    \label{fig: subfigures-3}
\end{figure}

Figs \ref{fig: subfigures-3}ab show $R_{\mathrm{TA}}$ versus the variance $\sigma^2$ of degree distributions for random regular graphs and perturbed ones with $N=6300$ nodes and the average degree $\langle k \rangle=4$ or 6.
$R_{\mathrm{TA}}$ is numerically calculated by the Newman-Ziff algorithm \cite{NZ-algo}.
In both Figs \ref{fig: subfigures-3}AB, $R_{\mathrm{TA}}$ increases as the variance $\sigma^2$ decreases in both discrete and random perturbations (blue triangles and green hexagons).
These results indicate that the smaller variance of the degree distribution tends to be more robust against attacks.
Furthermore, the random regular graphs (red circles) have the highest robustness than bimodal and modified networks.
Remember that the combinations of degrees for bimodal networks are comprehensive.
Therefore, it is strongly suggested that random regular graphs have the optimal robustness against attacks.

\begin{figure}[htp]
    \begin{minipage}[b]{0.53\textwidth}
        \includegraphics[width=\linewidth]{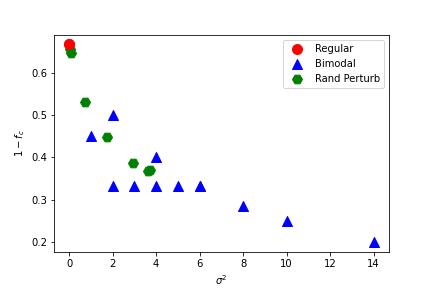}
        \caption*{(A) For $\langle k \rangle=4$.}
    \end{minipage}
    \begin{minipage}[b]{0.53\textwidth}
        \includegraphics[width=\linewidth]{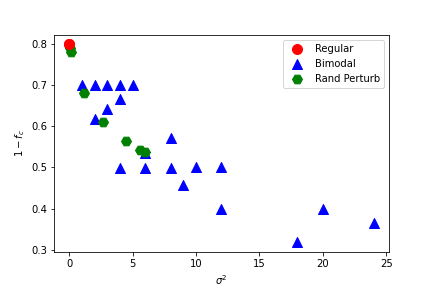}
        \caption*{(B) For $\langle k \rangle=6$.}
    \end{minipage}
    \caption{The percolation thresholds $1-f_c$ versus the variances $\sigma^2$ of degree distributions. The results for the average degree \textbf{(A)} $\langle k \rangle=4$, and \textbf{(B)} $\langle k \rangle=6$. In both figures, $1-f_c$ becomes higher as the variance decreases. Random regular graphs (red points) are the most robust. Note that a higher $1-f_c$ means more robust. }
    \label{fig: subfigures-4}
\end{figure}

For the theoretically estimated percolation threshold $1-f_c$, similar results are obtained for the robustness index $R_{\mathrm{TA}}$.
Figures \ref{fig: subfigures-4}ab show the percolation threshold $1-f_c$ versus the variance of degree distributions for random regular graphs and perturbed ones with $N=6300$ nodes and the average degree $\langle k \rangle=4$ or 6.
In both Figs \ref{fig: subfigures-4}AB, $1-f_c$ tends to increase as the variances $\sigma^2$ decrease for both discrete and random perturbations (blue triangles and green hexagons).  
Furthermore, the random regular graphs (red circles) have the highest robustness. 
Thus, random regular graphs have the optimal robustness against attacks.

\begin{figure}[htp]
    \begin{minipage}[b]{0.5\textwidth}
        \includegraphics[width=\linewidth]{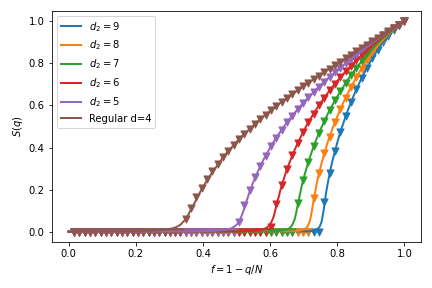}
        \caption*{(A) For $d_1=2$.}
    \end{minipage}  
    \begin{minipage}[b]{0.5\textwidth}
        \includegraphics[width=\linewidth]{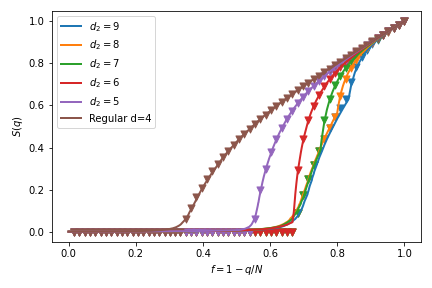}
        \caption*{(B) For $d_1=3$.}
    \end{minipage}
    \caption{The ratio of the largest connected components versus the fraction of remaining nodes on bimodal networks with $N=6300$ and $\langle k \rangle=4$. The results are shown for \textbf{(A)} $d_1=2$ and \textbf{(B)} $d_1=3$. }
    \label{fig: subfigures-5}
\end{figure}

In the bimodal networks (blue triangles in Figs \ref{fig: subfigures-3} or \ref{fig: subfigures-4}), 
there are similar values of $R_{\mathrm{TA}}$ or $1-f_c$ for different variances $\sigma^2$, e.g. $R_{\mathrm{TA}}$ takes about 0.25 for $\sigma^2=2$ and 4 in Fig \ref{fig: subfigures-4}A.
To investigate a relation between the robustness and the variances in bimodal networks, we show the ratio $S(q)$ of the largest connected components against attacks.
Fig \ref{fig: subfigures-5} show the ratio $S(q)$ in bimodal networks with $N=6300$ nodes and the average degree $\langle k \rangle=4$. 
The inverse triangles show the analytical results by Eq. (\ref{s_is_here}), while the solid lines show the results by numerical simulation using the Newman-Ziff algorithm \cite{NZ-algo}.
In Fig \ref{fig: subfigures-5}, $R_{\mathrm{TA}}$ is the area under the curve with respect to horizontal axis. 
Fig \ref{fig: subfigures-5}A for $d_1=2$ shows that $R_{\mathrm{TA}}$ decreases as $d_2$ increases.
On the other hand, Fig \ref{fig: subfigures-5}B for $d_1=3$ shows that $R_{\mathrm{TA}}$ decreases as $d_2$ increases, although the decrease of $R_{\mathrm{TA}}$ becomes smaller for $d_2>6$.
For example, there is a small difference between orange ($d_2=8$) and blue ($d_2=9$) lines.
In particular, in Fig \ref{fig: subfigures-5}B, the lines for $d_2>6$ are not smooth and like multi-step. 
Such a phenomenon is observed for $d_1=3$ close to the average degree $\langle k \rangle=4$.
However, the reason for this is not well understood.
Similar results are obtained for $\langle k \rangle=6$.

\section{Conclusion}
In this study, we find optimally robust networks against targeted attacks in varying degree distributions.
In considering that a network tends to be more robust with the smaller variance of degree distributions, random regular graphs with the minimum variance are predicted to be optimally robust.
Remember that it is insufficient to determine whether random regular graphs are optimally robust.
We clarify the optimal robustness in varying degree distributions by comparing the robustness of random regular graphs and their comprehensive discrete or random perturbations which includes several modalities of degrees.
By comparing the robustness index and percolation threshold on them, we find that random regular graphs are the highest robust in all bimodal and modified networks.
Our results show that random regular graphs have the optimal robustness against attacks in varying degree distributions.

\section*{Acknowledgements}
  This research is supported in part by JSPS KAKENHI Grant Number JP.21H03425.

\bibliographystyle{unsrt}  
\bibliography{mchujyo}
\end{document}